\documentclass[twocolumn,showpacs,preprintnumbers,amsmath,amssymb]{revtex4}
\reversemarginpar
\usepackage{graphicx}

\usepackage{bm}
\usepackage{amsfonts}
\usepackage{amssymb}
\usepackage{latexsym}
\usepackage{dcolumn}

\preprint{APS/123-QED}

\newcommand{\commentout}[1]{}
\newcommand{\nwc}{\newcommand}
\nwc{\rer}{r_{\eta,\rho}}
\nwc{\rinf}{r_{\eta,\infty}}

\nwc{\xvec}{{\vec{\bx}}}
\nwc{\kvec}{{\vec{\bk}}}
\newcommand{\lt}{\left}
\newcommand{\rt}{\right}

\newcommand{\ks}{\omega}

\newcommand{\bx}{\mathbf x}
\nwc{\mm}{\mathbf m}
\newcommand{\br}{\mathbf r}

\nwc{\bS}{\mathbf S}
\newcommand{\bp}{\mathbf p}
\newcommand{\by}{\mathbf y}
\nwc{\bX}{\mathbf X}
\nwc{\bY}{\mathbf Y}
\nwc{\bh}{\mathbf h}

\newcommand{\bw}{\mathbf w}

\newcommand{\bH}{\mathbf H}
\nwc{\bQ}{\mathbf Q}
\nwc{\bI}{\mathbf I}

\nwc{\nwt}{\newtheorem}
\nwt{assumption}{Assumption}

\nwc{\bal}{\begin{align}}
\nwc{\beq}{\begin{equation}}
\nwc{\ben}{\begin{equation*}}
\nwc{\bea}{\begin{eqnarray}}
\nwc{\beqa}{\begin{eqnarray}}
\nwc{\bean}{\begin{eqnarray*}}
\nwc{\beqn}{\begin{eqnarray*}}
\nwc{\beqast}{\begin{eqnarray*}}

\nwc{\eal}{\end{align}}
\nwc{\eeq}{\end{equation}}
\nwc{\een}{\end{equation*}}
\nwc{\eea}{\end{eqnarray}}
\nwc{\eeqa}{\end{eqnarray}}
\nwc{\eean}{\end{eqnarray*}}
\nwc{\eeqn}{\end{eqnarray*}}
\nwc{\eeqast}{\end{eqnarray*}}

\nwc{\tx}{\tilde{\bx}}
\nwc{\tp}{\tilde{\bp}}
\nwc{\tr}{\tilde{\br}}
\nwc{\tw}{\tilde{\bw}}
\nwc{\ep}{\varepsilon}
\nwc{\ept}{\epsilon}
\nwc{\vrho}{\varrho}
\nwc{\orho}{\bar\varrho}
\nwc{\ou}{\bar u}
\nwc{\vpsi}{\varpsi}
\nwc{\lamb}{\lambda}
\nwc{\wep}{W^\ep}

\nwc{\nn}{\nonumber}
\nwc{\mf}{\mathbf}
\nwc{\mb}{\mathbf}
\nwc{\ml}{\mathcal}

\nwc{\IA}{\mathbb{A}} 
\nwc{\IB}{\mathbb{B}}
\nwc{\IC}{\mathbb{C}} 
\nwc{\ID}{\mathbb{D}} 
\nwc{\IM}{\mathbb{M}} 
\nwc{\IP}{\mathbb{P}} 
\nwc{\II}{\mathbb{I}} 
\nwc{\IE}{\mathbb{E}} 
\nwc{\IF}{\mathbb{F}} 
\nwc{\IG}{\mathbb{G}} 
\nwc{\IN}{\mathbb{N}} 
\nwc{\IQ}{\mathbb{Q}} 
\nwc{\IR}{\mathbb{R}} 
\nwc{\IT}{\mathbb{T}} 
\nwc{\IZ}{\mathbb{Z}} 

\def\det{\mathop{\rm det}\nolimits}

\nwc{\cE}{{\ml E}}
\nwc{\cI}{{\ml I}}
\nwc{\cP}{{\ml P}}
\nwc{\cL}{{\ml L}}
\nwc{\cR}{{\ml R}}
\nwc{\cV}{{\ml V}}
\nwc{\cW}{{\ml W}}
\nwc{\cT}{{\ml T}}
\nwc{\crV}{{\ml V}_{(\delta,\rho)}}
\nwc{\cC}{{\ml C}}
\nwc{\cA}{{\ml A}}
\nwc{\cS}{{\ml S}}
\nwc{\cK}{{\ml K}}
\nwc{\cB}{{\ml B}}
\nwc{\cD}{{\ml D}}
\nwc{\cF}{{\ml F}}
\nwc{\cM}{{\ml M}}
\nwc{\cN}{{\ml N}}
\nwc{\cG}{{\ml G}}
\nwc{\cH}{{\ml H}}
\nwc{\bk}{{\mb k}}
\nwc{\cQ}{{\ml Q}}
\nwc{\cO}{{\ml O}}
\nwc{\cJ}{{\ml J}}
\nwc{\sir}{{\sf SIR}}
\nwc{\snr}{{\sf SNR}}
\nwc{\sinr}{{\sf SINR}}

\nwc{\mint}{{\int\cdot\int}}

\begin{document}

\title{Time Reversal Communication in Multi-Path  Fading Channels with Pinholes}

\author{Albert Fannjiang}
 \email{
  cafannjiang@ucdavis.edu}
 \affiliation{
Department of Mathematics,
University of California, Davis, CA 95616-8633}
\begin{abstract}
The paper presents an analysis of the time reversal 
in multi-path Rayleigh-fading channels with $N$
inputs (transmitters) and $M$ outputs (receivers).
 The main issues addressed 
are the condition of statistical stability, the rate of
information transfer and the effect of pinholes.
The stability condition is proved to be 
 $MC\ll N_{\rm eff}B$ for broadband channels and
 $M\ll N_{\rm eff}$ for narrowband channels where $C$ is the symbol rate,
 $B$ is the bandwidth and 
$N_{\rm eff}$ is the {\em effective} number of
transmitters. It is shown
that when the number of layers, $n-1$, is relatively low
compared to the logarithm of numbers of pinholes $N_{\rm eff}$  is given by
$n^{-1}$ times the harmonic mean of the number of transmitters and
the numbers of pinholes at all layers. 
On the other hand, when the number of layers is relatively
large the effective number of pinholes diminishes
exponentially. 
The energy efficiency is  shown to be optimal 
when the power supply is set to the noise level times
$BN_{\rm eff}$ and  that the maximal information rate is roughly $BN_{\rm eff}$ when the stability condition  is violated. 

\end{abstract}

\pacs{46.65.+g, 43.20.+g, 42.68.Ay}

\maketitle
\section{Introduction}
Time reversal (TR) of waves has received
great attention in recent years and been extensively studied 
for electromagnetic \cite{Ch}, \cite{LF}, \cite{Yanik} as well as acoustic propagation (see \cite{Fink} and the references therein). A striking effect of time reversal
in randomly inhomogeneous media is the superresolution
of refocal signals \cite{BPZ}, \cite{tire-phys}  which implies low probability
of intercept and holds high potential in technological
applications such as 
communications \cite{DTF}, \cite{RJD},  \cite{EK}, \cite{KK}.

An issue prior to superresolution, however, is
the condition of stability, namely
 how many antennas and how much bandwidth 
one needs 
to achieve statistical
stability in TR so that the received signals are nearly
deterministic, independent of the channel statistics?
 In this note
I answer this question for multi-path Rayleigh fading
channels,  with
multiple inputs and multiple outputs (MIMO),  commonly used in wireless communication literature, see, e.g. \cite{Paul}
and I analyze the effect of multi-layer pinholes
in Section~\ref{pin}. 

  In the MIMO-TR communication scheme \cite{DTF}, \cite{pulsa-pnas}, the $M$
well-separated receivers  first send a pilot signal to
the $N$-element TRA which then uses the time-reversed
version of the received signals to modulate the data symbols and retransmit them
back to the
receivers.
 One of the main results obtained here is that
the time reversal process is statistically stable when
\bea
\label{stab}
MC&\ll N_{\rm eff}B,&\hbox{for broadband channels}\\
M&\ll N_{\rm eff},&\hbox{for narrowband channels}
\label{stab2}
\eea
 where $C(\leq 2B)$ is the symbol rate,
 $B$ is the bandwidth and 
$N_{\rm eff}$ is the {\em effective} number of
transmitters. In the presence of $(n-1)$-layer pinholes, 
I show that the  effective number of
transmitters is asymptotically equal to
$n^{-1}$ times the harmonic mean of
the number of transmitters and the numbers
of pinholes of all layers when these numbers
are greater  than  $2^n$. That is, the multi-layer pinholes have
a screening effect on time-reversal transmission, reducing the effective number of  the time-reversal elements by at least a factor of $n$. When the last condition is violated, the situation is even worse:
$N_{\rm eff}$ diminishes exponentially fast as
the number of layers increases, resulting
in a tough environment to perform  time reversal of good quality.

The LHS of (\ref{stab}) is the number of degrees of freedom
per unit time
in the constellation of intended data-streams while
the RHS of (\ref{stab}) is roughly 
the number of degrees of freedom per unit time in the channel state information (CSI)
received by TRA from the pilot signals.  The latter
has to be larger than the former in order to reverse 
the random scrambling by the channel and achieve
deterministic outputs. The stability condition $N\gg 1$
for narrow-band channels
or $B\gg \beta_c$ (the coherence bandwidth) for
broadband channels, when $M$ is  small  and the pinholes 
are absent,  have been previously discussed in
\cite{BPZ}, \cite{DTF1}, \cite{DTF2}, \cite{DTF}, \cite{LF}.

In Section~\ref{pin},  I take into account the effect of noise
 and analyze the information
rate in TR communication in noisy channels.
I demonstrate
a tradeoff between stability and spectral efficiency:
the maximal information rate $R\sim BN_{\rm eff}$ is achieved when
the inequality in (\ref{stab})-(\ref{stab2}) is reversed. Also,
for a given level of noise the energy efficiency is
optimized  when the power level is set to
the noise level times $N_{\rm eff} B$.

\section{TR-MIMO communication}
 First let us review the MIMO-TR communication scheme
 as described in \cite{pulsa-pnas}. 

\begin{figure}
\begin{center}
\includegraphics[width=6.5cm, totalheight=4cm]{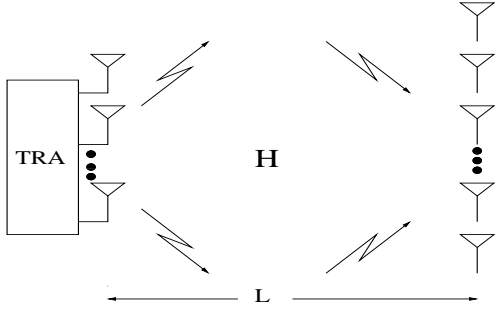}
\end{center}
\caption{MIMO Channel
 }
 \end{figure}

 The $M$
receivers  located at $\by_j, j=1,...,M$ first send a pilot signal $\int e^{i {\ks t}}g(\ks)$$ d\ks\delta(\bx_j-\by_i)$ to
the $N$-element TRA located at $\bx_i, i=1,...,N$ which then use the time-reversed
version of the received signal $\int e^{i {\ks t}}g(\ks) H(\by_j, \bx_i;\ks)d\ks$ to encode a stream of symbols and retransmit them
back to the
receivers. Here $H$ is the transfer function of
 the propagation channel at the frequency $\ks$
 from point $\by$ to $\bx$ and  $g^2(\ks)$ is the power
density at $\ks$.  Let $\bH(\ks)=[H_{ij}(\ks)]$ be the transfer matrix between
 the transmitters and receivers where
 $
 H_{ij}(\ks)=H(\bx_i,\by_j;\ks).$  The reciprocity implies
 that $\bH(\ks)$ is symmetric and  the relation
 $\bH^*(\ks)=\bH(-\ks)$ where $*$ stands for complex conjugation.   Let us assume
that $g$ is a smooth and rapidly decaying
function with effective support of size $B$ such as
 $g^2(\ks)=(2\pi)^{-1/2}\exp{(-\frac{|\ks-\ks_0|^2}{2B^2})}$.  Naturally  
 the relative bandwidth $B/\ks_0$ is less than unity. 
 We have chosen the time unit such that
the speed of propagation is one. 

Let us assume that the separation $L$ between the TRA and the receivers is much larger than the spacing within the TRA-elements
and the receivers.  The
 the signal vector $\bS=(S_j)$ arriving at the  receiver with delay $L+t$ is then given by \cite{BPZ}, \cite{DLF} 
  \beqa
 \nn
{S_j(t)}
&= &\sum_{l=1}^W\sum_{i=1}^M m_i(\tau_l)\int  e^{-i{\ks}(t-\tau_l)}g(\ks)\\
&&\times\sum_{k=1}^NH_{jk}(\omega)H^*_{ik}(\ks) d\ks
\label{mr}
\eeqa
where $m_j(\tau_l), l=1,...,W\leq\infty$ is a stream of symbols intended for the $j$-th
receiver transmitted at times $\tau_l=l\tau,\tau>0 $.  
In vector notation, we have $\bS
=\sum_{l=1}^W \int e^{-i\ks(t-\tau_l)}g(\ks)\bH\bH^\dagger(\ks)\mm(\tau_l)
  d\ks$ where
 $\bH^\dagger $ is the conjugate transpose of $\bH$ and
 $\mm(\tau_l)=(m_j(\tau_l))$.
Let us note that while all the TRA-elements are coordinated
and 
synchronized the receivers do not have the knowledge about
the channel
 and can not coordinate in decoding the total signals
received. As a consequence, the co-channel interference
from multiple users can  be a serious problem \cite{Paul}.
An advantage of the time reversal scheme is the possibility to use
the (statistical) stability property to achieve
the following asymptotic
\[
\int  e^{-i{\ks}(t-\tau_l)}g(\ks)\sum_{k=1}^NH_{jk}(\omega)H^*_{ik}(\ks) d\ks
\sim \delta_{ij}\delta(t-\tau_l)
\]
so that $S_j(t)\sim \sum_{l=1}^Wm_j(\tau_l) \delta(t-\tau_l)$ and each receiver receives the intended symbols without  interference.

\section{Statistical stability}
 One of
the main goals of the present note is to characterize  the stability regime
for the important channel model  of multi-path Rayleigh fading in which $H_{ij}$ are
 i.i.d. $CN(0,1)$, the zero-mean, variance-one circularly symmetric complex-Gaussian random variables.
 For simplicity, I assume that $|m_i(\tau_l)|=\mu, $$
 \forall i,l$. The multi-path Rayleigh fading after proper normalization is a simplified model for richly scattering
 environment
 when the spacing  within the transmitters
 and  receivers is larger than 
 the coherence length $\ell_c$ of the channel.
In general, the coherence length  is inversely proportional to
 the angular spread \cite{Paul} and sometimes
 can be computed explicitly in terms of physical properties
 of the channel \cite{pulsa-pnas}. 
For diffuse waves the coherence length is known to be on the scale of
wavelength \cite{Sh}, \cite{SS}.

 Let us calculate
 the mean and the variance of the signals with respect to
 the ensemble of the channel. We use $\IE$ to denote
 the channel ensemble average. Consider the
 quantity $\bH\bH^\dagger\mm$. By the Gaussian rule for
 the calculation of moments we have
 \beqa
 \IE \bH\bH^\dagger\mm&=&N\mm\\
  \IE \big|\big(\bH(\ks)\bH^\dagger(\ks)\mm\big)_j\big|^2&=&
  N^2|m_j|^2+N \sum_{i=1}^M|m_i|^2
  \label{a.4}
  \eeqa
  so that
 $\IE\bS=NB \mm \sum_{l=1}^We^{-i\ks_0(t-\tau_l)}e^{-B^2(t-\tau_l)^2}$.   Let $\tau\geq (2B)^{-1}$ so that the  summation in $\IE \bS$ is $B$-uniformly bounded as $W\to\infty$. 
 
We measure the statistical stability of
 the signals by  the normalized variance  of the signals at the receivers
 \[
 \cV_j(\tau_n)=\frac{V_j(\tau_n)}{|\IE S_j|^2(\tau_n)},
 \,\, V_j(\tau_n)\equiv \IE |S_j|^2(\tau_n)-|\IE S_j(\tau_n)|^2,
 \]
 $\forall j, n$
 and say that the signals are stable when $\cV_j(\tau_n)\to 0, \forall j,n$.
 
 Let $\beta_c$ be  the coherence bandwidth of the channel such that 
 \beqn
\lefteqn{\IE\Big\{\big(\bH(\omega)\bH^\dagger(\omega)\mm\big)_j\big(\bH(\omega')\bH^\dagger(\omega')\mm\big)^*_j\Big\}}\\
   &\approx&
 \left\{\begin{array}{ll}
N^2|m_j|^2+N \sum_{i=1}^M|m_i|^2,& \quad |\omega-\omega'|\leq\beta_c/2,\\
 0,&\quad |\omega-\omega'|\gg\beta_c.
\end{array}
 \right.
 \eeqn
The coherence bandwidth $\beta_c$ is  inversely proportional
 to the delay spread  and hence 
 the delay-spread-bandwidth product (DSB) is roughly $B\beta_c^{-1}$ \cite{2f-whn}, \cite{pulsa-pnas}, \cite{Paul}.  
 In the diffusion approximation $\beta_c$ is given 
 by the Thouless frequency $D_BL^{-2}$ where
 $D_B$ is the Boltzmann diffusion constant, equal to the energy transport velocity times the transport mean free path, and $L$ the distance
 of propagation  \cite{LT}, \cite{Sheng}. 
 
The {\em broadband, frequency-selective} (BBFS) channel
is naturally defined as having a large DSP, i.e. $B\beta^{-1}_c\gg 1$.
Since $B<\ks_0$, $\ks\in [\ks_0-B/2, \ks_0+B/2]$ and $-\ks$ are separated by more than $\beta_c$ in a broadband frequency-selective channel. The same holds for
a {\em  narrow-band, frequency-non-selective} (NBFN) channel defined as $ B\leq\beta_c\ll \ks_0$.

Consider the NBFN case first. We have 
  \bean
 V_j(t)\approx N B^2\sum_{i=1}^M|m_i|^2\big|\sum_{l=1}^We^{i\ks_0\tau_l}e^{-B^2(t-\tau_l)^2/2}\big|^2
  \eean
In view of (\ref{a.4})  the stability condition $N\gg M$ for NBFN channels then follows easily. The main focus of the paper, however, is  the BBFS channels for which  we have instead
  \bea
 \label{1.3}V_j(t)
 &\approx& N B\sum_{i=1}^M|m_i|^2\sum_{l}^W\frac{2\sin{\frac{\beta_c}{2}(t-\tau_{l})}}{t-\tau_l}\\
 &&\times
\sum_{l'=1}^We^{-i\ks_0(l'-l)\tau}e^{-B^2(l-l')^2\tau^2/2}.\nn
  \eea
  Several observations are in order. First, 
due to $\tau\geq (2 B)^{-1}$  the summation over $l'$ in (\ref{1.3})
is convergent as $W\to \infty$ uniformly in $l$ and $B$.
Second, the summation over $l$ is also convergent as
$W\to\infty$ with the effective number of terms
$\sim B\beta_c^{-1}$.
As a result, it suffices to consider the case $ W=O( B \beta_c^{-1})$ or equivalently $|\tau_1-\tau_W|=O(\beta_c^{-1})$
for which we have  the estimate $
V_j\sim NBC\sum_{i=1}^M|m_i|^2$
 where $C$ is the number of symbols per unit time in {\em each}
 data-stream.
 It then follows  that
 $\cV_j\to 0$ if and only if
 $NB\gg MC$ for BBFS channels. The transition to the
 condition $N\gg M$ for NBFN channels takes place
 when $B\sim C$, i.e. $\tau\sim B^{-1}$.
 
 Since $NB$ is the number of degrees of freedom in the channel
 state information collected 
 at the TRA per unit time and $MC$ is the number
 of degrees of freedom in the ensemble of messages per unit time
 the stability condition $NB\gg MC$ can be interpreted as saying
 that in order to recover the deterministic messages, independent
 of the channel ensemble, and thus reverse the random scrambling
by the channel  the former must be much larger
 than the latter.

A detailed,  rigorous analysis 
of the MIMO-TR 
channel modeled 
by a stochastic Schr\"odinger equation, in the parabolic
approximation,  with
a random potential is given in \cite{pulsa-pnas}.
 \section{Rate of information transfer}
In this section we discuss the information rate for a
memoryless channel based on the multi-path Rayleigh-fading 
transfer matrix defined above \cite{Tel}, \cite{FG}. In a memoryless channel an independent realization of the random
 transfer matrix is drawn  after  each delay spread.  
 Let us assume as usual that 
 the noise at the each receiver is additive-white-Gaussian-noise (AWGN) and that the input vector is multivariate Gaussian
 and that the channel, the receiver noise and the input
 are independent of one another. 
 
 According to Shannon's theorem \cite{CT} the information (in nats) that a symbol
 can convey on average 
  is 
  $2^{-1}\ln{(1+\sinr)}$ where $\sinr$, the signal-to-interference-noise ratio at each receiver,  is given
  by the harmonic mean of the $\sir$, the signal-to-interference ratio and $\snr$, the signal-to-noise ratio,  i.e.
  $\sinr=(\sir^{-1}+\snr^{-1})^{-1}$. This formulation assumes
  that the interference is approximately AWGN, like the noise
  at the receivers. For the multi-path Rayleigh fading
  channel considered here the interference statistics is strictly
  speaking high-degree $\chi$-square for which Gaussian
  statistics is a reasonable approximation for our purpose of deriving a rough estimate
  of the tradeoff between the stability and  information rate.
  
  According to the preceding analysis $\sir\sim NB/(MC)$ in the
  case of BBFS channels and $\sir\sim N/M$ in the case
  of NBFN channels, independent of the power constraint.
 Let us set  the covariance matrix of the receiver noises
 to be $\nu\bI_M$, the $M$-dim identity. Suppose the average transmission  power is constrained to 
  $P$ is constrained and all
  the transmit and receive  antennas are identical. We have $\snr=P/(\nu MC)$. 
  In the case
  of BBFS channels, the $\sinr$ is roughly
  \beq
  \label{sinr}
  \sinr\sim \Big(\frac{MC}{NB}+\frac{\nu MC}{P}\Big)^{-1}
  \eeq
  and, in the case of NBFN channels, it is
    \beq
  \label{sinr2}
  \sinr\sim \Big(\frac{M}{N}+\frac{\nu MC}{P}\Big)^{-1}
  \eeq
  The optimal $\sinr$ is obtained by setting $N\sim{P(\nu \max{(B,C)})^{-1}}$ which is roughly $P/(vB)$ since $C\leq 2B$. The information rate $R$, given roughly by $2^{-1}MC\ln{(1+\sinr)}$, achieves the maximum roughly equal to
  $ P/\nu$ (in nat) at $MC\gg NB$ (for BBFS) or $N\gg M$ (for NBFN). That is,  the channel capacity is linearly  proportional to the power
  and is achieved at the expense of statistical stability
  of signals.
 
 Consider the thermal noise power $\nu=k_B TB$ where
 $k_B$ is the Boltzmann's constant and $T$ the temperature.
 Then the above result implies that the energy cost per nat is $ P/R\sim k_B T$ 
 (with a constant close to unity)
 which is an extension to the TR-MIMO Rayleigh fading  channel of the classical result  derived for a simple SISO channel \cite{Pei}
 (see also \cite{Le}). 
 \commentout{
  Choosing the Gaussian
 input is to maximize the mutual information of
 the input and output which
 in this case is given by 
 \beq
 \cI(X;Y)=\log{\det{\lt(\bI_M+\bQ\bH^*\bH\rt)}}
 =\log{\det{\lt(\bI_M+\bH^*\bH\bQ\rt)}}.
 \eeq
 see, e.g. \cite{Tel}. Since $\bH^*\bH$ and $\bH\bH^*$ share
 the same set of nonzero eigenvalues, 
 
 The
 resulting channel capacity is given by \cite{Tel}, \cite{FG}
 \[
 \cC=\int_B \log_2{\det{(\bI_N+\frac{\rho}{M}\bH\bH^*(\ks))}}
 d\ks
 \]
 where $B$ is the frequency band of the signals and
 $\rho$ is the average SNR (independent of $M$)
 at each receive antenna.
 }

 \section{Pinhole effect}
 \label{pin}
 \begin{figure}
\begin{center}
\includegraphics[angle=270,width=6.5cm, totalheight=4cm]{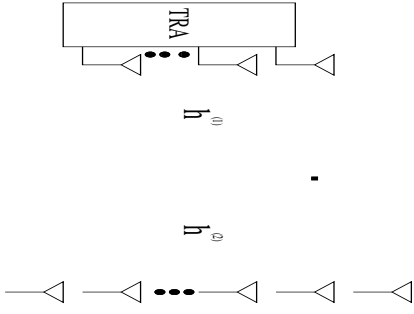}
\end{center}
\caption{Single-layer pinholes
 }
 \end{figure}

 Pinholes are degenerate channels  that can occur 
 in a wide family of channels, outdoor as well as indoor, see
 Figs. 2 and 3. While preserving the co-channel decorrelation,  pinholes have has been shown to severely  limit
 the degrees of freedom and reduce the channel capacity \cite{CFV}, \cite{GP}, \cite{CFG}. In this section, I present an analysis
 of the pinhole effect on  TR in Rayleigh fading channels to demonstrate similar effects on stability and information rate.
 
 First, let us consider the simplest case of single-layer pinholes
 as illustrated in Fig. 2. Let $\bh^{(1)}(\omega)$ be the $N\times K$ transfer matrix  from the TRA to the pinholes and $\bh^{(2)}(\omega)$
 the $K\times M$ transfer matrix from the pinhole
 to the $M$ receivers at frequency $\omega$. The combined channel can be
 described by $\bH(\omega)=\bh^{(2)}(\omega)\bh^{(1)}(\omega)=[h^{(2)}_{ik}(\omega) h^{(1)}_{kj}(\omega)]$
 in which $ h^{(1)}_{kj}$ and   $h^{(2)}_{ij}$ are assumed
 to be i.i.d. $CN(0, \sigma_1)$ and $CN(0,\sigma_2)$,
 respectively. 
 
At frequency $\omega$,
 the mean signals received are given by $
 \IE \bH\bH^\dagger \mm=NK\sigma_1\sigma_2\mm$ and  the variance
 of the signal at frequency $\ks$ received at receiver $i$ 
 is given by 
 \bean
&&\sum_{k=1}^K\IE|h^{(2)}_{ik}|^2\sum_{n=1}^M  \IE|h^{(2)}_{nk}|^2|m_n|^2\big(\sum_{j=1}^N\IE|h^{(1)}_{kj}|^2\big)^2\\
&&+\sum_{k=1}^K\IE|h^{(2)}_{ik}|^2\sum_{l=1}^K\sum_{n=1}^M\IE|h^{(2)}_{nl}|^2|m_n|^2 \sum_{j=1}^N\IE|h^{(1)}_{kj}|^2\IE|h^{(1)}_{lj}|^2\\
&&+\sum_{k=1}^K\big(\IE|h^{(2)}_{ik}|^2\big)^2|m_i|^2\big(\sum_{j=1}^N\IE|h^{(1)}_{kj}|^2\IE|h^{(1)}_{kj}|^2\big)\\
&&\approx KN(MN+MK+1)\sigma^2_1\sigma_2^2|\mu|^2.
 \eean
 Taking into account  the temporal aspect of the signal as before
 we obtain 
 the normalized variance of the signal to the leading order  
 ($N,K\gg 1$) for BBFS channels
 \[
\cV_j\approx {MC}B^{-1}\big(N^{-1}+K^{-1}\big),\quad\forall j 
\]
and
for NBFN channels
 \[
\cV_j\approx {M}\big(N^{-1}+K^{-1}\big),\quad\forall j.
\]
The result indicates  that there
is an {\em effective} number of TRA-elements given by
$
N_{\rm eff}=NK\big(N+K\big)^{-1},
$
namely  one half of the harmonic mean of $N$ and $K$, so that
$\cV_j\approx MC B^{-1}N_{\rm eff}^{-1}$ for BBFS channels
and $\cV_j\approx MN_{\rm eff}^{-1}$ for NBFN channels . The confirms
the intuition that pinholes are choke-points that
reduce the effective number of TRA-elements. 

The previous case {\em without }
 pinholes corresponds to the limiting case $K\to\infty$. 
For a fixed  $K$, however,  the previous benefit of having large number of TRA elements  
 ($N\gg 1$) disappears.  The multiple antennas in TRA 
 are essentially screened out by the pinholes and
 the effective number  of TRA-elements becomes $K$.
 
 \begin{figure}
\begin{center}
\includegraphics[width=6.5cm, totalheight=4cm]{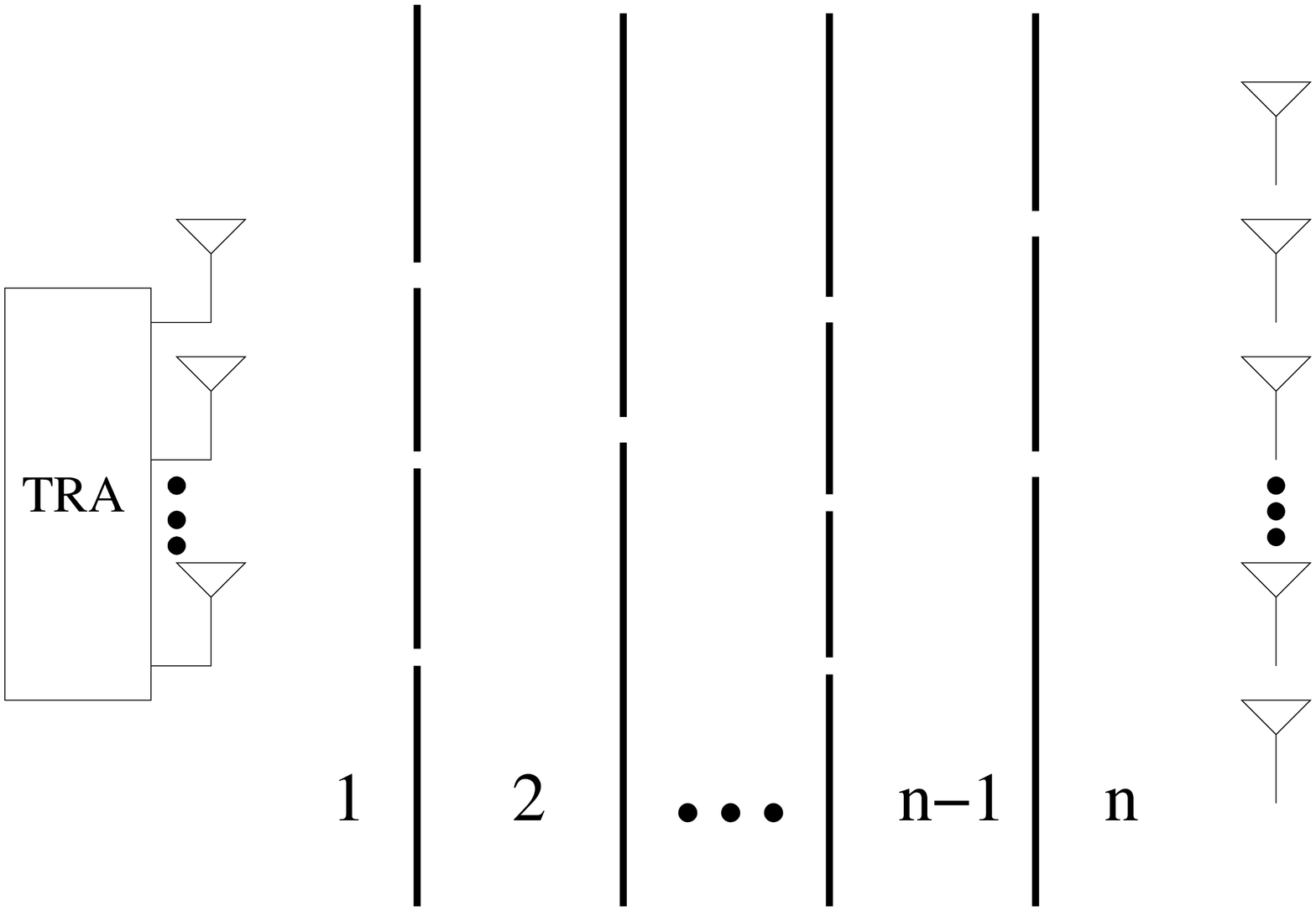}
\end{center}
\caption{Multi-layer pinholes
 }
 \end{figure}

 The same analysis can be applied to 
 channels with $(n-1)$ layers of pinholes such as illustrated in Fig. 3.
 Let $K_k, k=1,..n-1$ be the number of $k$-th layer pinholes.   
 Let $\bh^{(k)}$ be the transfer matrix for the $k$-th stage channel whose entries are i.i.d. zero-mean, variance-$\sigma_k$ Gaussian r.v.s. and let the transfer matrices of different
 stages be independent of one  another.
 
I now show that to the leading order ($N, K_1,..,K_{n-1}\gg 2^n$)
the normalized variance  of the signal  is given by
$\cV\approx MCB^{-1}N_{\rm eff}^{-1}$ where
the effective number of TRA-element $N_{\rm eff}$ is given by
\[
N_{\rm eff}=\Big(N^{-1}+N^{-1}_{\rm p}\Big)^{-1},\quad N_{\rm p}=\Big(\sum_{j=1}^{n-1} K_j^{-1}\Big)^{-1};
\]
namely $n^{-1}$ times the harmonic mean 
of $N, K_1, \cdots, K_{n-1}$.  The {\em effective} number of {\em pinholes}
$N_{\rm p}$ is always larger than $N_{\rm eff}$ and
this again confirms the limiting nature of the pinholes.
I sketch the proof below. 

The calculation of the mean
is straightforward: $\IE \bH\bH^\dagger\mm=NK_1\cdots  K_{n-1}$. Let us analyze the second moment of entry $a$
\bea
\label{gauss}\nn
\lefteqn{\IE \Big(\bH\bH^\dagger\mm\Big)_a\Big(\bH\bH^\dagger\mm\Big)_a^*}\\
\nn&=&\IE\sum_{i_1,\cdots i_n\atop j_2,\cdots  j_{n+1}}
h_{a i_n}^{(n)}h_{i_n, i_{n-1}}^{(n-1)}\cdots h^{(2)}_{i_3, i_2}h^{(1)}_{i_2, i_1}h^{(1)*}_{j_2, i_1}h^{(2)*}_{j_3, j_2}\\
&&\times \cdots h^{(n-1)*}_{j_n, j_{n-1}}
h^{(n)*}_{j_{n+1}, j_n}m_{j_{n+1}}\sum_{i'_1,\cdots i'_n\atop j'_2,\cdots  j'_{n+1}}
h_{a i'_n}^{(n)*}h_{i'_n, i'_{n-1}}^{(n-1)*}\nn\\
&&\times\cdots h^{(2)*}_{i'_3, i'_2}h^{(1)*}_{i'_2, i'_1}h^{(1)}_{j'_2, i'_1}h^{(2)}_{j'_3, j'_2}\cdots h^{(n-1)}_{j'_n, j'_{n-1}}
h^{(n)}_{j'_{n+1}, j_n}m^*_{j'_{n+1}}.\nn
\eea
I claim that according to the Gaussian rule the leading order terms of the second moment correspond to the
{\em simple} graphs in which the arcs, connecting (un)primed indices to (un)primed indices, are nested and are bound by
the ladder edges, connecting unprimed indices to  primed indices.
This includes the graph corresponding to $\big|\IE\bH\bH^\dagger\mm\big|^2$ which has
no ladder edges. A simple graph is illustrated in Fig. 4.

\begin{figure}
\begin{center}
\includegraphics[width=6.5cm, totalheight=3cm]{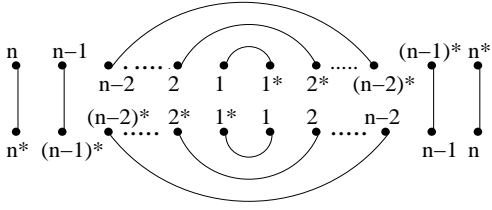}
\end{center}
\caption{Example of a simple graph
 }
 \end{figure}

The observation  is  proved by induction. When a new layer of pinholes, described by $\bh^{(n+1)}$,
is added, the number of graphs is doubled: one half of them
contain the ladder edges connecting  $h^{(n+1)}_{ai_{n+1}}$
to $h^{(n+1)*}_{ai'_{n+1}}$ and
$h^{(n+1)*}_{j_{n+2}, j_{n+1}}$ to 
$h^{(n+1)}_{j'_{n+2}, j'_{n+1}}$ while the other half
contain the arcs connecting $h^{(n+1)}_{ai_{n+1}}$
to $h^{(n+1)*}_{j_{n+2}, j_{n+1}}$ and $h^{(n+1)*}_{ai'_{n+1}}$
to $h^{(n+1)}_{j'_{n+2}, j'_{n+1}}$.  A moment of reflection
reveals that the new pair of arcs 
impose {\em one} more constraint on the ranges of the indices
 than the new pair of ladder edges  and yield a linear factor of $K_{n+1}$ or $M$ to
the $n$-th order graphs while the new pair of ladder edges yield
a quadratic factor $MK_{n+1}$ to the graph corresponding to $\big|\IE\bH\bH^\dagger\mm\big|^2$ and $K_{n+1}^2$
to the rest of the $n$-th order simple graphs. 
Note that
the structure of simple graphs is not changed by the new ladder edges and the total number of $n$-th order simple graphs
is exactly $n+1$.  

Collecting the terms corresponding to the simple graphs 
 we have
\bean
&&\mu^2NM\prod_{i=1}^n\sigma_i^2
 \prod_{j=1}^{n-1}K_j\\
 &&\times\big(\prod_{k=1}^{n-1} K_{k}
 +N\sum_{i=1}^{n-1}K_1\cdots \widehat K_i\cdots K_{n-1}\big)
 \eean
  where $\widehat K_i$ means that $K_i$ is absent in the
 product. Dividing it by $N^2\prod_{i=1}^{n-1}K_i^2$ and accounting for the temporal aspect of transmission we obtain
 the claimed result.

To calculate the $\sinr$ we can substitute $N_{\rm eff}$ for $N$ in (\ref{sinr}). Even with unlimited power supply  one 
would do best by choosing the optimal power level $P\sim B\nu N_{\rm eff}$; higher power level is wasteful. The maximal information rate is roughly $ BN_{\rm eff}$ which is still bounded  by $BN_{\rm p}$.

\section{Conclusions}

I have analyzed the time reversal of propagation
in multi-path Rayleigh-fading MIMO-channels
with or without pinholes. The focus of the analysis is
the stability condition, the multiplexing gain and their
tradeoff.
The main results can be summarized as (i)
that the stability holds when $MC\ll N_{\rm eff}B$ for
BBFS channels and $M\ll N_{\rm eff}$ for
NBFN channels
where $N_{\rm eff}$ is the effective number of
TRA-elements and (ii) that 
the maximal information rate $BN_{\rm eff}$ is attained
when the power supply is set to the noise level times
$BN_{\rm eff}$  and when
the stability condition is reversed.
They are a significant extension of 
the previously discussed  conditions for stability
\cite{BPZ}, \cite{DTF1}, \cite{DTF2}, \cite{DTF}.

I have computed the {\em effective} number
of TRA-element $N_{\rm eff}$ under the
condition that  the numbers
of TRA-elements and the pinholes of each layer   are much
greater than $2^n$, with $n-1$ being the number of layers. 
In this extreme case the effective number of TRA-elements
is asymptotically equal to 
$n^{-1}$ times the harmonic mean of TRA-elements and the numbers of pinholes at all $n-1$ layers.
In the opposite case  the graph analysis shows that the normalized
variance  of signal grows exponentially with
the number of pinhole layers and consequently the rate
of 
information transmission diminishes exponentially fast.
In other words, a long chain of independently fluctuating media
separated by a series of screens of sparse pinholes is detrimental
to time reversal communication. 

The estimate $NB$ for the TR information rate in the absence of pinholes   should be contrasted with the finding  
 in
\cite{FG}, \cite{Tel}, \cite{Mou}, \cite{SM} 
that the capacity with CSI at
the receiver with $M$ receive antennas (but not at the 
$N$ transmit antennas) scales like $ B\min{(M,N)}\ln{\snr}$ at high {\sf SNR}. 
Note that 
  their result does not include the interference due to non-cooperating
  multiuser receivers as is done here and is
  for  narrow-band signals with $C=2B$. 

\acknowledgments 
 This research is supported in part by  DARPA Grant N00014-02-1-0603,
 NSF grant DMS 0306659.

\end{document}